\documentclass[%
 preprint,
 amsmath,amssymb,
]{revtex4-1}

\usepackage{graphicx}
\usepackage{dcolumn}
\usepackage{bm}
\usepackage{hyperref}
\usepackage{amsmath}

\usepackage[utf8]{inputenc}
\usepackage{booktabs}

\usepackage{color}
\definecolor{redcolor}{rgb}{1.0,0.,0.}

\begin{document}

\preprint{APS/123-QED}

\title{Comparing the impact of subfields in scientific journals}

\author{Xiomara S. Q. Chacón}
\affiliation{%
Institute of Mathematics and Computer Science, University of S\~{a}o Paulo,
S\~{a}o Carlos, SP,  Brazil
}%

\author{Thiago C. Silva}

\affiliation{Universidade Cat\'olica de Bras\'ilia, Distrito Federal, Brazil}
\affiliation{Department of Computing and Mathematics, Faculty of Philosophy, Sciences, and Literatures in Ribeirão Preto, Universidade de S\~ao Paulo, S\~ao Paulo, Brazil}



\author{Diego R. Amancio}
\email{diego@icmc.usp.br}
\affiliation{%
Institute of Mathematics and Computer Science, Department of Computer Science, University of S\~{a}o Paulo,
S\~{a}o Carlos, SP,  Brazil
}%

\date{\today}

\begin{abstract}
The impact factor has been extensively used in the last years to assess journals visibility and prestige. While the impact factor is useful to compare journals, the specificities of subfields visibility in journals are overlooked whenever visibility is measured only at the journal level. In this paper, we analyze the subfields visibility in a subset of over 450,000 Physics papers. We show that the visibility of subfields is not regular in the considered dataset. In particular years, the variability in subfields impact factor in a journal reached 75\% of the average subfields impact factor.
We also found that the difference of subfields visibility in the same journal can be even higher than the difference of visibility between different journals. Our results show that subfields impact is an important factor accounting for journals visibility.
%
\end{abstract}

\maketitle


\section{Introduction}

In recent years, the visibility of scientific papers, authors, journals and conferences have been used as an important feature to quantify research relevance and impact~\cite{waltman2016review}. The number of citations has been an important quantity to gauge the visibility and quality, and for this reason, many research impact measurements have been devised based on citation counts~\cite{redner1998popular}. At the author level, for example, the h-index has been widely used as a proxy to scientific relevance~\cite{bornmann2007we}, despite the many criticisms~\cite{egghe2006improvement}.

Citations also plays an important role  in evaluating journals research output. The prestige of scientific journals is oftentimes measured via citation counts, among other factors~\cite{glanzel2002journal}. One important citation index for journals is the \emph{impact factor} (IF), which essentially gives the average number of citations received by papers in a journal in the last 2 years. In many cases, researchers use journals impact factor (and other journal attributes) to identify the most relevant venue to disseminate their research. While the impact factor has been a disseminated index to measure visibility and relevance, it has been mostly used at the journal level~\cite{alberts2013impact}. In this sense, the specificities of subfields in journals have been mostly overlooked. This means that in a high impact journal, some subfields might have a lower visibility. Conversely, in a low impact journal, particular subfields might be more visible than the journal as a whole.

In this paper, we investigate the behavior of subfields visibility in scientific journals. We analyze whether there is a significant difference of visibility for different subfields in the same venue. Our analysis was performed in 450,000 Physics papers published by the \emph{American Physical Society}. Some interesting results have been found. First we found that there is a considerable variability in subfields impact in some cases. The variability in the subfields impact factor can reach up to 75\% of the average impact factor of subfields in the \emph{Physical Review Letters} (PRL) journal. In recent years, the variability reached $50\%$ of PRL impact factor. We also found that the difference in  the visibility of subfields of the same journal might be higher than difference in the visibility of journals. This results suggests that not only the venue is an important factor to predict papers visibility, but also the subfield inside the journal.

The study of subfields impact is important because it provides a granular view of journals impact. This information could be used by authors to provide a more informed decision on the choice of journals to publish. The analysis of subfields impact might also be useful to understand the dynamics of journals visibility along time. A more refined visibility information in journals could also assist editors in identifying promising subfields or subfields that are no longer representative in terms of visibility.


%

\section{Related works} \label{sec:related}

Many factors have been found to affect papers visibility. The total number of citations might not be influenced only by the inherent quality and innovation~\cite{bornmann2015does}. The total number of citations received in the recent past might be an indicative of how many citations a paper will receive in the future~\cite{amancio2012three}. This process is referred to as preferential attachment in the network science field~\cite{jeong2003measuring,newman2001clustering}.

The journal in which the scientific paper is published is important to establish the correct audience for the paper, which may affect the future number of citations~\cite{lariviere2010impact,mckiernan2019use,bornmann2015does}. Another important feature that could affect papers visibility is the prestige of the journal. Oftentimes, prestige is gauged by visibility measurements -- such as the impact factor, CiteScore, eigenfactor and influence score~\cite{abramo2010citations}.

Prestige at the author level plays an important role in the success of papers. Renowned authors are naturally more visible and therefore tend to receive more citations~\cite{amjad2017standing}. The preferential attachment is also a relevant factor driving the dynamics of authors' citations. A recent model showed that a more reliable description of the citation curve of authors should take into consideration only the citations accrued by authors in the last 12-24 months~\cite{recency}.

Some additional factors have also been found to have an effect on the visibility of papers. This includes interdisciplinarity of the subject being approached, the number of tables, figures, references and some textual factors, such as the title length~\cite{leydesdorff2007betweenness,rostami2014effect,leydesdorff2011indicators,onodera2015factors,amancio2012using}. While many of these studies focus on the visibility of journals, papers and major fields, here we perform a visibility analysis at the subfield level. More specifically, we analyze the impact of different subfields inside journals.

\section{Methodology}

\subsection{Dataset and PACS classification}

The \emph{Physics and Astronomy Classification Scheme} (PACS) is a hierarchical classification of Physics and Astronomy scientific papers. A PACS code comprises 3 elements: a pair of two digits separated by a dot. The digits are followed by two characters (letters or positive and negative symbols). In the first part of the code, the first digit represents the main category of the paper and the second digit is the subfield inside the field specified by the first digit. The last characters in the code provide an even more specific characterization of subfields. For instance, the PACS code $05.45.-a$ refers to the following classification: ``$0$'' represents the ``\emph{General}'' field, ``$05$'' denotes ``\emph{Statistical physics, thermodynamics, and nonlinear dynamical systems}''. Finally, the last part ``-a'' represents the ``\emph{Nonlinear dynamics and chaos}'' subfield.
In our work we focused our analysis at the third hierarchical level, which corresponds to the code ``$05.45$'' in the previous example. This is an intermediary hierarchical level that allows us to analyze subfields that are neither too general nor too specific. A list of all subfields mentioned in this paper is available in the Supplementary Information.

The dataset we used for the current paper is the dataset provided by the \emph{American Physical Society} (APS). This dataset provides citations for the APS journals: \emph{Physical Review Letters} (PRL), \emph{Review of Modern Physics} (RMP) and \emph{Physical Review A}-\emph{E} (PRA-E). We obtained the citation data for over 450,000 papers published in APS journals between 1983 and 2016. While our analysis used only the citation data, journal name, PACS code and publication date, the APS dataset also provides additional information, such as DOI, title, authors names and affiliations. For each journal and year, we considered a subfield relevant if at least 50 articles were published in that field in the last two years.


\subsection{Comparing groups of papers} \label{sec:sucInd}

In this paper we compare the impact of subfields. The subfields might belong to the same journals or to different journals. The difference in the visibility of subfields was  computed using the so-called \emph{citation success index}~\cite{milojevic2017citation}. We decided to use this measurement because it provides a clear interpretation of the differences in the impact factor (and citation distribution) between any two groups of papers. The success index $\mathcal{S}$ is designed to quantitatively compare the success of two journals, but the same concept can be extended to compare any set of papers.  Given two set of papers, the reference ($r$) and target ($t$) sets,  the success index $\mathcal{S}_{tr}$ is defined as the probability that a randomly drawn article from the target group will receive more citations than an article drawn from the reference group. $\mathcal{S}_{tr}$ can be computed directly from the citation distribution of $t$ and $r$~\cite{milojevic2017citation}:
\begin{equation} \label{eq:suc_ind}
    \mathcal{S}_{tr} = \sum_{c=0}^\infty \Bigg{(} {P}_t(c) + \frac{p_t(c)}{2} \Bigg{)} p_r(c),
\end{equation}
where ${P}_t(c)$ is the fraction of papers in $t$ with more than $c$ citations, and $p_r(c)$ is the fraction of papers in $r$ that received $c$ citations.

A relationship between the citation success index and the impact factor can be derived from the definition of impact factor and the definition of the success index in equation \ref{eq:suc_ind}. The following equation can also be used to compute the citation success index:
\begin{equation} \label{eq:siif}
    \mathcal{S}_{tr} = \frac{f_0}{2} + \frac{ 1 - f_0/2 }{ 1 + q \rho^{-k} }
\end{equation}
where $\rho = I_t / I_r$, is the ratio between impact factors of sets $t$ and $r$, respectively; $k=1.23$, $q$ is a normalization factor and $f_0$ is the rate of uncited papers in $r$. It has been shown that $f_0$ can be described by the following logistic function~\cite{milojevic2017citation}, which allows us to obtain $q$ from
\begin{equation}
    q = \frac{1}{1-f_0}.
\end{equation}
The computation of $\mathcal{S}_{tr}$ from equation \ref{eq:siif} can be simplified whenever $f_0$ is low (which typically occurs when $I_r > 10$) or $I_t > I_r$. In this case,
\begin{equation}
    \mathcal{S}_{tr} = \frac{ 1  }{ 1 + \rho^{-k} }.
\end{equation}

\section{Results and Discussion} \label{sec:results}

In Section \ref{sec:st}, we analyze some subfields statistics in journals. In Section \ref{sec:2}, we compare the visibility of subfields in the \emph{same} journal. In Section \ref{sec:3}, subfields of \emph{distinct} journals are compared.

\subsection{Subfields statistics} \label{sec:st}

\begin{figure}[h]
\centering
{%
\includegraphics[scale=0.9]{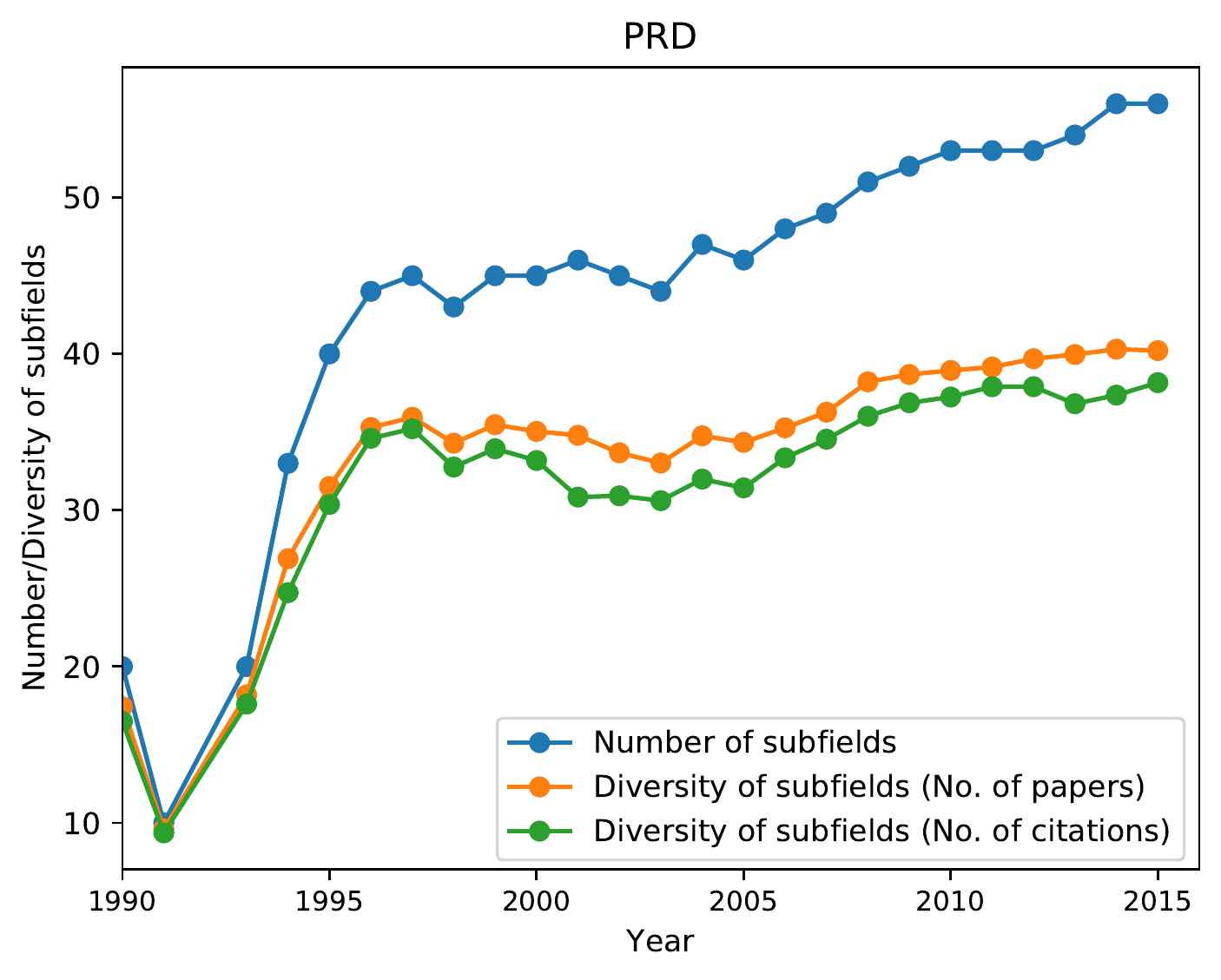}
\caption{\label{fig:prddiversity}Evolution in the number of PRD subfields (blue curve). The orange and green curves represent the diversity of subfields considering the number of papers and number of citations in PRD subfields, respectively.}
}
\end{figure}

We start our study by analyzing the evolution in the number of relevant subfields of journals. Overall, in most of the considered journals, the number of subfields increases along the recent years (see blue curve in Figure \ref{fig:prddiversity}). We also measured the number of subfields by considering the heterogeneity in the number of articles published in each subfield. To do so, we computed the diversity of subfields, a measurement that has been used in network science and other fields~\cite{tuomisto2010consistent,correa2017patterns,de2017knowledge,jost2006entropy}.
According to the diversity index, if all fields have the same size, the diversity of subfields corresponds to the total number of subfields. Conversely, if the only a few subfields have most of the published articles, the diversity of subfields will be much smaller than the total number of subfields. In the diversity index, such a heterogeneity is measured via entropy~\cite{correa2017patterns}.

The diversity of subfields measured in terms of the number of articles published for each subfield is shown in  Figure \ref{fig:prddiversity} (orange curve). Note that there is a heterogeneity in the number of papers in each subfield, since the of subfields diversity  is much smaller than the total number of fields. In 2015, almost 60 subfields were found; however, the diversity points to \emph{effectively} only 40 subfields. We also computed the diversity of subfields considering the number of citations received by subfields (rather than the number of published papers). For the PRD journal, Figure \ref{fig:prddiversity} shows that both diversity measurements are  similar. This suggests that of number of citations received by subfields
follows the number of published papers. 

A different scenario can be observed for both PRE and PRL journals in recent years (result not shown). While the diversity curves for the number of papers and citations follow the same behavior, the diversity of subfields considering the number of papers is higher than the diversity measured via citations. For instance, in 2015, 115 subfields were identified in PRL. In this same year, the diversity of subfields in terms of the number of papers was about 90 subfields. Conversely, the diversity of subfields measured in terms of citations was only about 78 subfields. This results suggests that some subfields are more cited than others and this difference cannot be explained only by subfields size in PRL. To further investigate the differences in subfields visibility, in the next section we compare subfields in the same journal using the citation success index.

\subsection{Comparing subfields in the same journal} \label{sec:2}

\begin{figure}[h]
\centering
{%
\includegraphics[scale=.85]{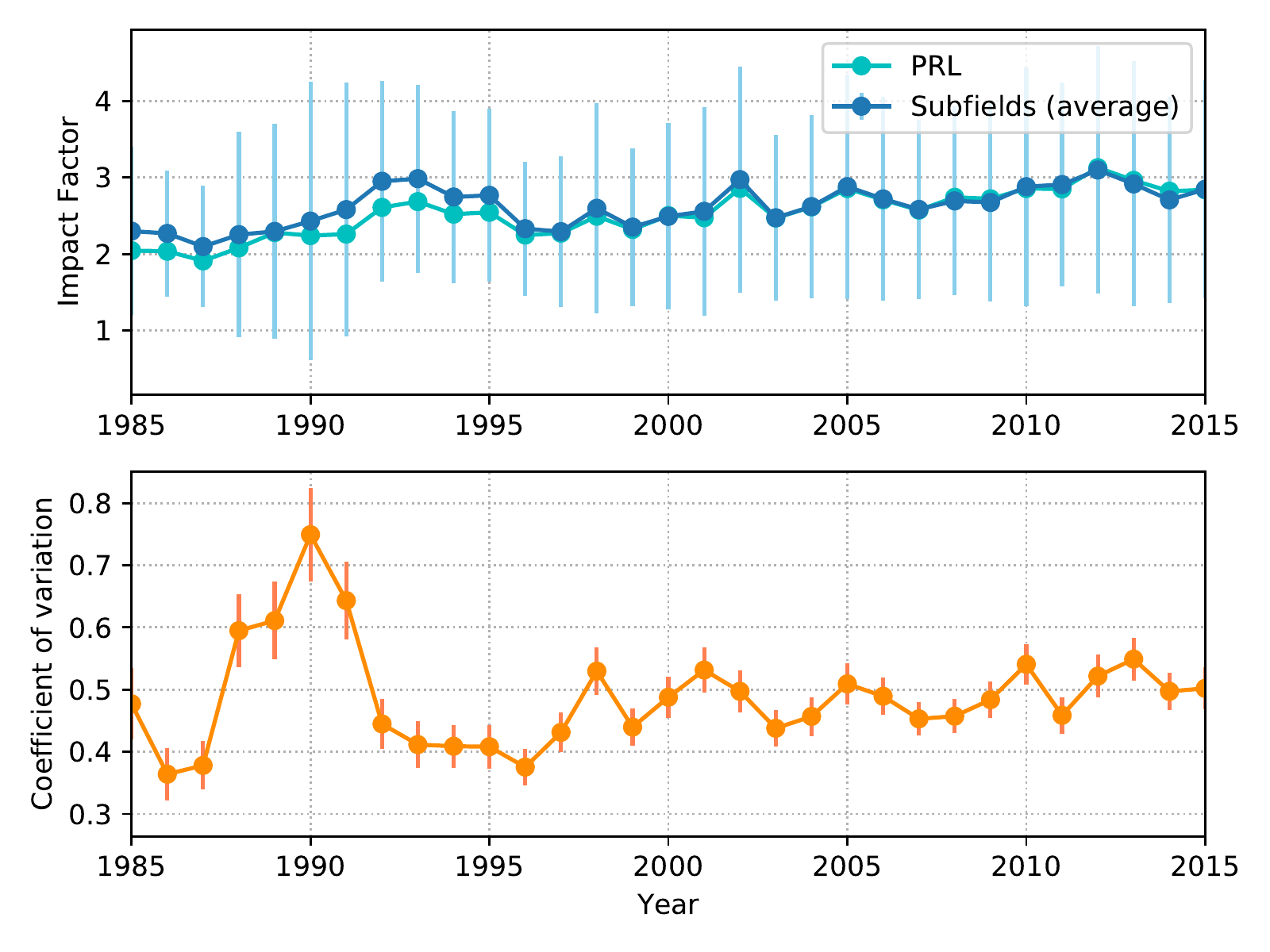}}
\caption{Evolution of the impact factor for subfields in the PRL journal. In (a), we show both average and standard deviation of subfields impact factor. In (b), we show the coefficient of variation of subfields impact factor.
}
\label{fig:prl1}
\end{figure}

In order to analyze how the  subfields visibility varies along time, we computed the yearly impact factor (IF) of each PACS code. In Figure \ref{fig:prl1}(a), we show the evolution of average impact factor of PRL subfields. We also show the evolution of PRL impact factor. Because we used the APS dataset, we are limited to the citations received by APS journals. As a consequence, impact factors might not be the same reported by \emph{Clarivate Analytics}~\footnote{\url{https://clarivate.com/}}. Similarly to other results in literature, this sampling does not affect the comparison of journals and subfields citation data~\cite{milojevic2017citation,recency}.

The average subfield IF is consistent with the journal IF. This happens for all considered journals. We however observe a variability of subfields impact along time, meaning that different subfields are more (or less) visible than the journal as a whole. Such a variability is evident when analyzing the coefficient of variation of subfields IF. Figure \ref{fig:prl1}(b) reveals that, in the most recent years, the typical deviation is roughly 50\% of the average IF. An even higher heterogeneity of subfields impact occurred in 1990. In that year, a typical deviation of 75\% of the average IF was observed. This result suggests that different subfields being published by PRL might have different visibility. Other APS journals display a similar behavior, however the coefficients of variation of subfields impact are typically below 0.50 (result not shown).

To further investigate how different is the impact of subfields inside a journal, we used the citation success index (see Section \ref{sec:sucInd}). This measurement provides a clearer interpretation regarding the difference of visibility (i.e. impact factor) between two subfields. In other words, given subfields $A$ and $B$, the success index $\mathcal{S}_{AB}$ comparing them gives the probability that a randomly drawn article from $A$ will be more cited than an article drawn $B$ (see Section \ref{sec:sucInd}).

\begin{figure}[h]
\centering
{%
\includegraphics[scale=1.5]{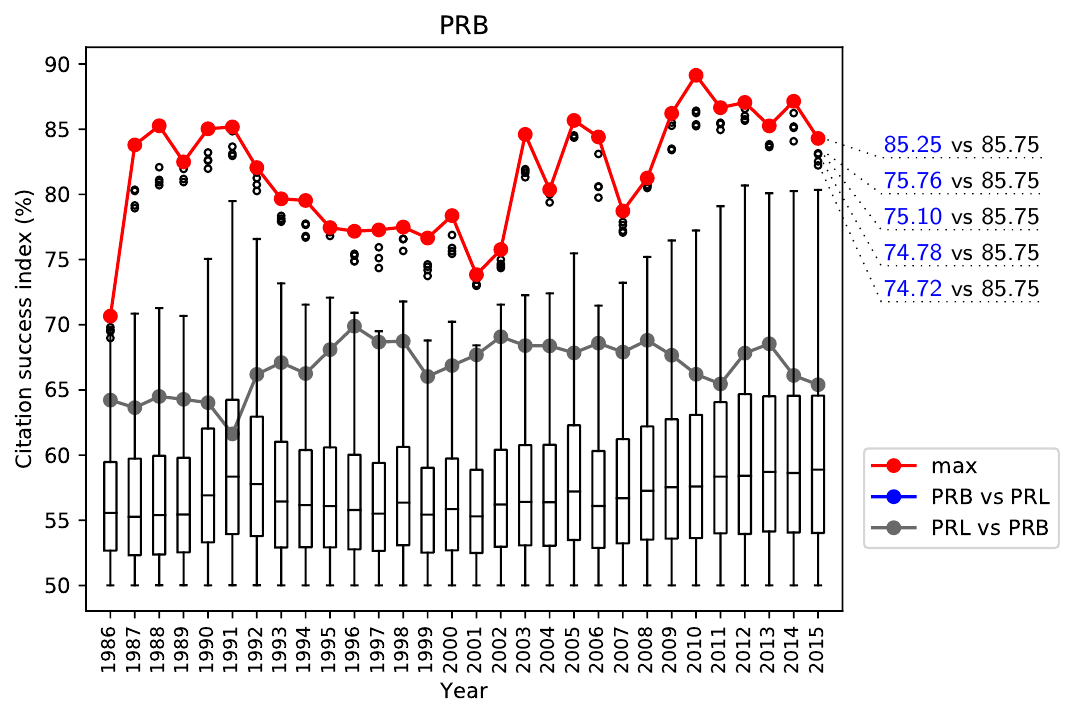}
\caption{\label{fig:fig1}Evolution of success index comparing the impact of all pairs of PRB subfields. The red curve represents the highest success index obtained for a given year. The gray curve represent the success index resulting from the comparison of PRL and PRB. Typically, the impact difference between PRL and PRB is more significant than the impact difference between PRB subfields. A list of PACS codes is shown in the Supplementary Information.}
}
\end{figure}

To understand the differences in visibility, for each journal, we measured the
the success index between all pairs of subfields in the same journal. Because $\mathcal{S}_{AB} + \mathcal{S}_{BA}= 100\%$, in our analysis, for each pair $A$ and $B$,
we only considered the maximum between $\mathcal{S}_{AB}$ and $\mathcal{S}_{BA}$. The results obtained for PRB is shown in Figure \ref{fig:fig1}. For this particular journal, we note that the median success index (comparing all pairs of subfields) varies between 55-57\%. This means that typically the impact factor of subfields in PRB are similar. However, in particular cases, some subfields are much more visible than others. The highest values of success index is highlighted in the red curve. For comparison purposes, we also show the success index obtained when comparing PRL and PRB (gray curve). Typically, the difference of visibility between PRL and PRB is more significant than the difference in visibility of PRB subfields. However, for \emph{particular} pairs of subfields, the difference of visibility between subfields is more significant than the difference of impact between journals (PRL and PRB). In 2015, the success index comparing PACS 85.25 and 85.75 reached almost 85\%, while the difference between PRB and PRL was roughly 65\%.

\begin{figure}[h]
\centering
{%
\includegraphics[scale=1.5]{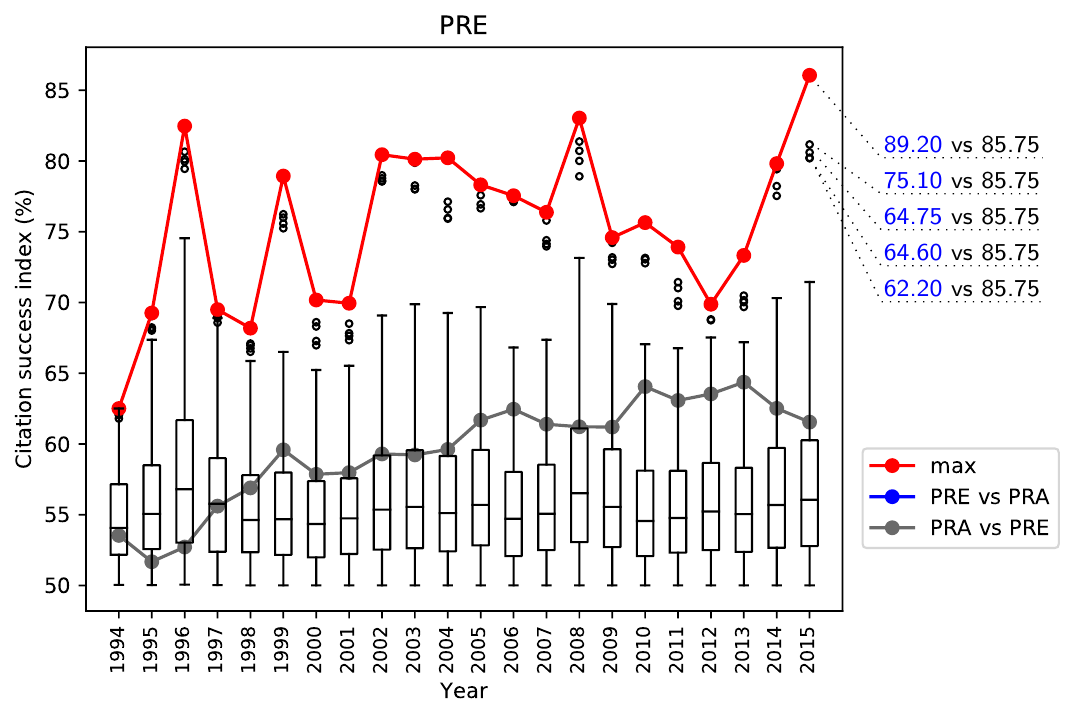}
\caption{\label{fig:figpre}Evolution of success index comparing the impact of all pairs of PRE subfields. The red curve represents the highest success index obtained for a given year. The gray curve represents the success index resulting from the comparison of PRA and PRE. A list of PACS codes is shown in the Supplementary Information.}
}
\end{figure}

In Figure \ref{fig:figpre}, we show the distribution of success indexes when comparing all pairs of PRE subfields (boxplots). In all considered years, most of the success indexes are below 60\%, revealing that there is no significant impact difference in PRE subfields for most of the considered pairs. However, as observed for PRB (see Figure \ref{fig:fig1}), some pairs of subfields display very distinct impact factors. In 2015, the comparison between PACS 89.20 and 85.75 yielded a success index higher than 85\%. This is much higher
In recent years, we can see that the typical internal impact difference is lower than the impact difference between PRA and PRE (see gray curve). Differently from Figure \ref{fig:fig1}, in terms of visibility, it seems the choice of subfield inside PRE is more important than the choice between PRA and PRE in particular years. This is clear e.g. in 1995 and 1996, when more than 75\% of all pairs of subfields were found to have a visibility difference higher than the one observed between PRA and PRE.

Figure \ref{fig:figpra} shows the distribution of success indexes comparing all pairs of PRA subfields. We also show the evolution of the success index comparing PRA and PRB.
This comparison is shown in two different colors: gray; meaning that the impact of PRB is higher then the impact of PRA; and blue, to represent the opposite. Unlike the previous analysis, we observe an interesting behavior after 1997. The difference in impact between PRA and PRB becomes very small, as revealed by success indexes typically below 55\%. At the same time, PRA subfields impact difference are typically larger than 55\%, once again meaning that the internal (subfield) visibility differences might be more important than the visibility differences observed between journals. In extreme cases, the success index comparing two PRA subfields reaches almost 90\%.
\begin{figure}[h]
\centering
{%
\includegraphics[scale=1.5]{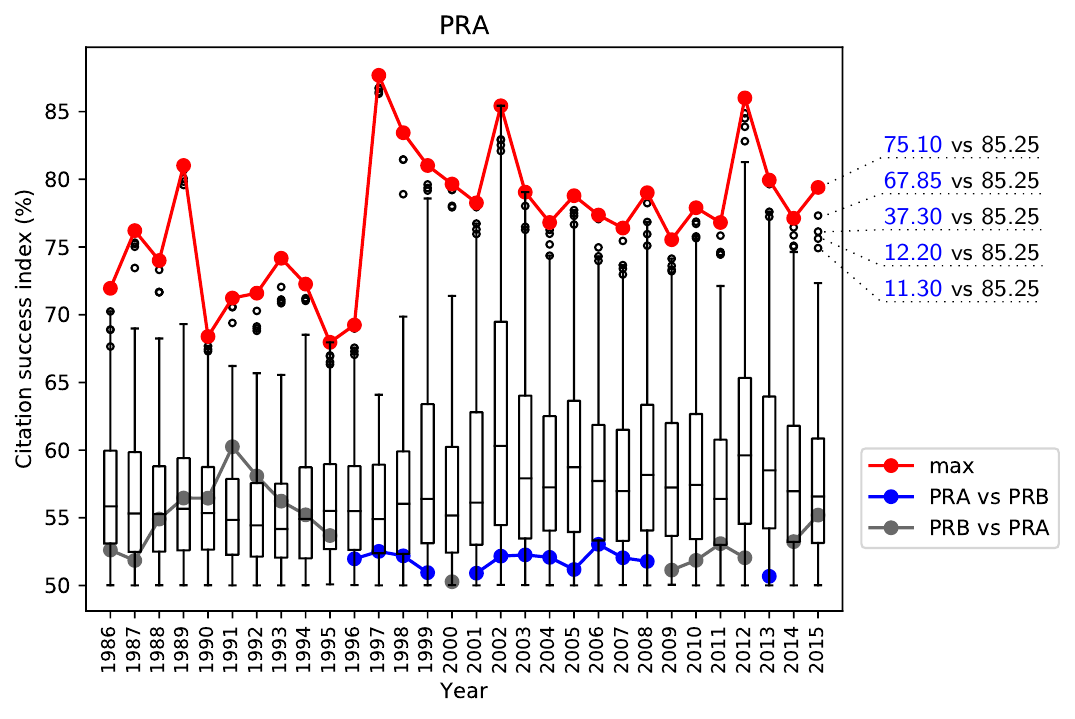}
\caption{\label{fig:figpra}Evolution of success index comparing the impact of all pairs of PRA subfields. The red curve represents the highest success index obtained for a given year. The gray and blue curves represent the success index resulting from the comparison of PRA and PRB impact. A list of PACS codes is shown in the Supplementary Information. }
}
\end{figure}

\subsection{Comparing subfields of different journals}  \label{sec:3}

\begin{figure}[h]
\centering
{%
\includegraphics[scale=1.5]{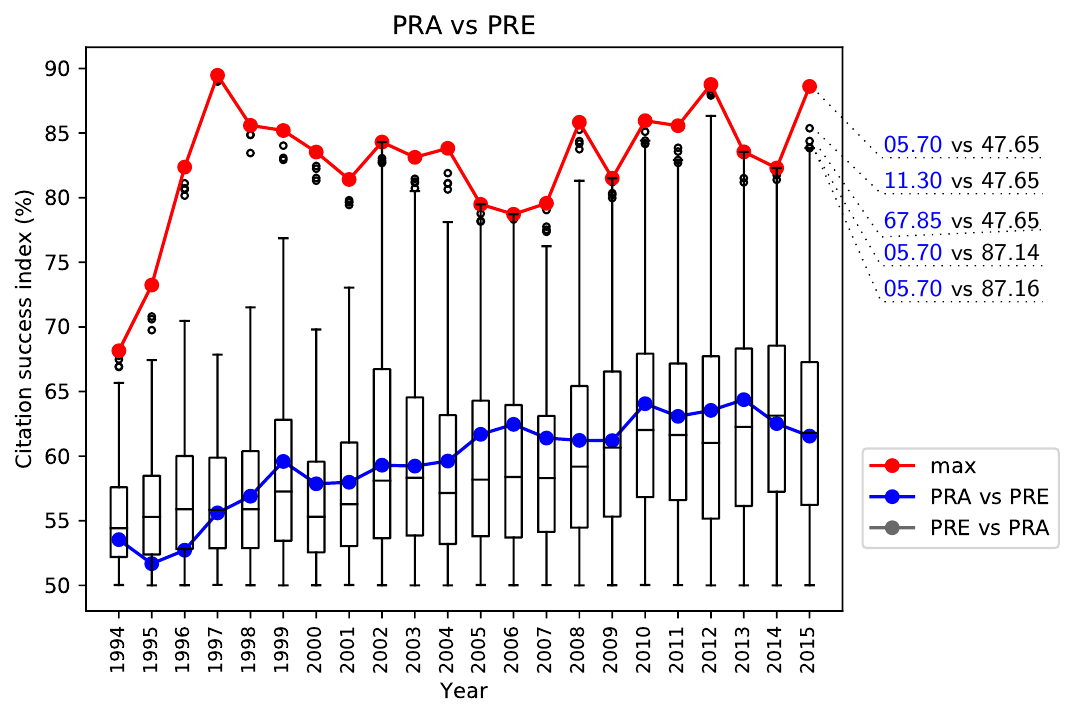}
\caption{\label{fig:figprapre}Evolution of success index comparing the impact of PRA and PRE subfields. The red curve represents the highest success index obtained for a given year. The gray and blue curves represent the success index resulting from the comparison of PRA and PRE impact. A list of PACS codes is shown in the Supplementary Information.}
}
\end{figure}

While in Figures \ref{fig:fig1}--\ref{fig:figpra} we focused our analysis on the comparison of subfields in the \emph{same} journal, we now compare the impact factor of subfields in different journals. The comparison between all pairs of PRA and PRE subfields is shown in Figure \ref{fig:figprapre}. One interesting pattern observed here is that the curve of success index comparing both journals follows the same behavior of the median comparing pairs of subfields. Therefore, in general, the visibility comparison of PRA and PRE is compatible with the comparison of the respective subfields visibility. However, particular subfields have very distinct visibility, as revealed by the dynamics of the red curve in Figure \ref{fig:figprapre}. Particularly, in 2015, a very large difference in visibility was found when comparing PACS 05.70 (from PRA) and 47.65 (from PRE). Note that such a difference in visibility is much higher than the typical visibility difference between PRA and PRE.

\begin{figure}[h]
\centering
{%
\includegraphics[scale=1.5]{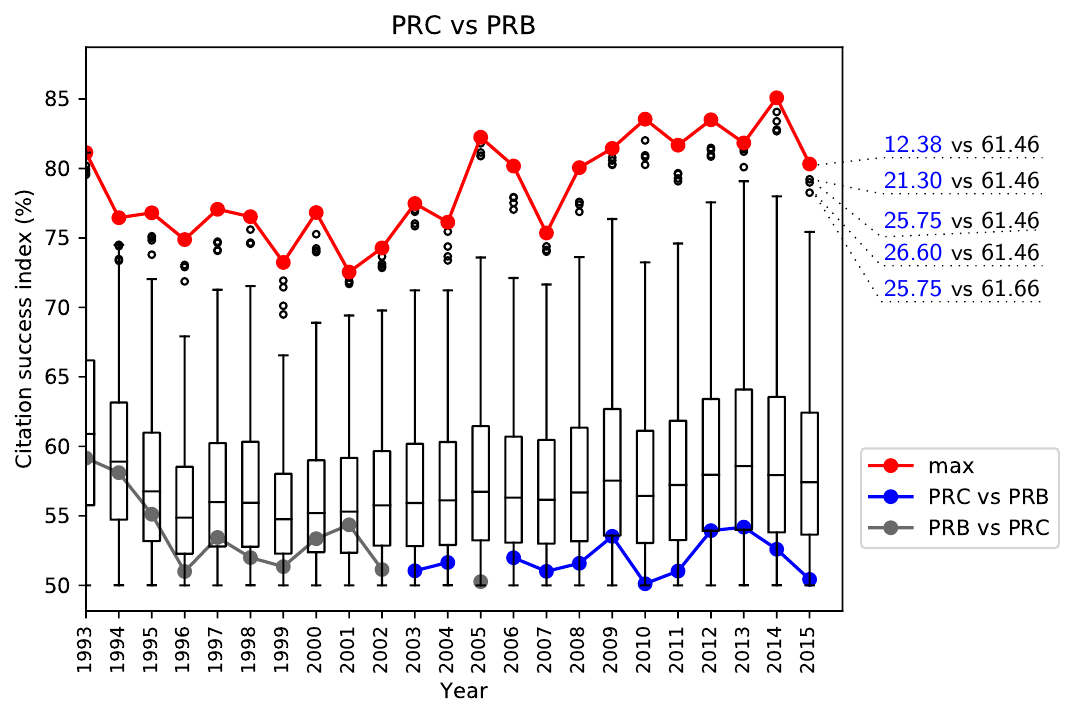}
\caption{\label{fig:figprcprb}Evolution of success index comparing the impact of PRC and PRB subfields. The red curve represents the highest success index obtained for a given year. The gray and blue curves represent the success index resulting from the comparison of PRA and PRE impact. A list of PAC codes is shown in the Supplementary Information.  }
}
\end{figure}

The relevance of comparing subfields (rather than journals) can be noted when comparing subfields from PRC and PRB journals, as shown in Figure \ref{fig:figprcprb}. From 1986 to 1994, the success index comparing PRB and PRC is compatible with the median of the success index comparing the respective subfields. However, from 2002 to 2015, it is clear that PRC and PRB have very similar values of impact factor, as revealed by values of success index very close to 50\% (see gray and blue curves). In this same period, however, the typical difference between subfields visibility was close to 60\%. Once again, for particular subfields, the citation success index reached values close to 85\%.  While in 2015 the journals have the same impact factor, the subfields represented by PACS 12.38 and 61.46 were found to have a difference in impact yielding  a success index close to 80\%.

All in all, the results presented in sections \ref{sec:2} and \ref{sec:3} revealed that subfields intra- and inter- journals might have very distinct visibility. This is an important result since it could be used as an additional information in research policies. From  scholars' perspective, the quantification of subfields visibility could assist researchers in their career path decisions. Along with other field attributes, it could be of interest to know beforehand the potential visibility of subfields before efforts are made to learn and produce knowledge in the field.

%

Another interesting conclusion arising from the obtained results concerns the comparison of journals. Our results show that in some cases the direct comparison of journals might not give all the information relevant. Two journals with similar impact factors might have very distinct visibility values when subfields are compared. This information could be used e.g. to improve predictions and evaluations that use the impact factor (in combination with other established research impact indexes).
%

\section{Conclusion} \label{sec:conc}

The analysis of subfields impact is relevant to provide  a more detailed information of the factors affecting journals visibility. In this paper we studied the variability of subfields impact in a subset of Physics journals published by the \emph{American Physical Society}. The identification of subfields was performed using the classification scheme provided by APS. One interesting result arising from our study is that the difference in the visibility of subfields in a same journal might be higher that the difference of visibility between journals.
Altogether, our results suggest that subfields visibility in journals might be not uniform and this information could be used to better understand the components affecting journals impact.

While we focused the subfields visibility analysis via PACS classification, this work could be extended by considering other notions of subfields~\cite{stasanew}. For example, subfields could be identified without any type of classification scheme. In particular, the identification of subfields could be performed via community detection in citation (or co-citation) networks~\cite{silva2016using}. Subfields could also be identified using co-occurrence word networks obtained from title and/or abstract~\cite{castro2019multiplex,amancio2015concentric,stella2019forma}. Collaboration networks could also be used to detect subfields~\cite{viana2013time}. Finally, we also intend to extend this study to analyze the variability of subfields visibility in other major fields.

\section*{Acknowledgments}

\noindent
DRA acknowledges financial support from São Paulo Research Foundation (FAPESP Grant no. 16/19069-9) and CNPq-Brazil (Grant no. 304026/2018-2). TCS gratefully acknowledges financial support from the CNPq foundation (Grant no. 408546/2018-2). XSQC acknowledges Capes-Brazil for sponsorship.

\newpage

\section*{Supplementary Information: Sublist of PACs}

This Supplementary Information lists the code of the subfields mentioned in this manuscript. The first two digits corresponds to the first hierarchical levels. In this study we analyzed subfields at the third hierarchical level. This hierarchical level comprises 3 digits.

\begin{enumerate}

\item \textbf{Code 05}: statistical physics, thermodynamics, and nonlinear dynamical systems, subfield: thermodynamics (05.70).

\item \textbf{Code 11}: General theory of fields and particles, with subfields: symmetry and conservation laws (11.30).

\item \textbf{Code 12}: specific theories and interaction models; particle systematics, with subfields: quantum electrodynamics (12.20), quantum chromodynamics (12.38).

\item \textbf{Code 21}: nmuclear structure, with subfield: nuclear forces (21.30).

\item \textbf{Code 25}: nuclear reactions: specific reactions, with subfields: relativistic heavy-ion collisions (25.75).

\item \textbf{Code 25}: nuclear astrophysics, with subfields: nuclear matter aspects of neutron stars (26.60).

\item \textbf{Code 37}: mechanical control of atoms, molecules, and ions, with subfields: atoms, molecules, and ions in cavities (37.30).

\item \textbf{Code 47}: fluid dynamics, with subfields: magneto-hydrodynamics and electro-hydrodynamics (47.65).

\item \textbf{Code 67}: quantum fluids and solids, with subfields: ultracold gases, trapped gases (67.85).

\item \textbf{Code 62}: Mechanical and acoustical properties of condensed matter, with subfield: mechanical properties of solids (62.20).

\item \textbf{Code 64}: equations of state, phase equilibria, and phase transitions, with subfields: general studies of phase transitions (64.60), phase equilibria (64.75).

\item \textbf{Code 67}: quantum fluids and solids, subfield: ultracold gases, trapped gases (67.85).

\item \textbf{Code 74}: superconductivity, with subfields: cuprate superconductors (74.72), superconducting films and low-dimensional structures (74.78).

\item \textbf{Code 75}: magnetic properties and materials, with subfields: general theory and models of magnetic ordering (75.10), spin transport effects (75.76).

\item \textbf{Code 85}: electronic and magnetic devices; microelectronics, with subfields: superconducting devices (85.25), magnetoelectronics; spintronics: devices exploiting spin polarized transport or integrated magnetic fields (85.75).

\item \textbf{Code 87}: biological and medical physics, with subfields: biomolecules: types (87.14), subcellular structure and processes (87.16).

\item \textbf{Code 89}: other areas of applied and interdisciplinary physics, subfield: interdisciplinary applications of physics (89.20).




\end{enumerate}

\newpage

\bibliographystyle{ieeetr}

\end{document}